\documentclass[aps,pra,twocolumn,groupedaddress,showpacs]{revtex4}

\begin{document}

\title{Elementary Thermodynamics of Trapped Particles}

\author{Martin Ligare}
\email[]{mligare@bucknell.edu}
\affiliation{Department of Physics, Bucknell University, Lewisburg, PA 17837}


\begin{abstract}
I develop simple thermodynamic relations for a collection of
noninteracting classical particles confined in a harmonic trap. The
volume of such a trap is not a good thermodynamic variable, so
conventional expressions of the first law of thermodynamics and the
ideal gas law must be modified.  I use the frequency of oscillations
about the minimum of the trap as an external parameter characterizing
the confinement, and derive elementary relations between particle
number, temperature, energy, oscillation frequency, and a generalized
pressure, that are analogous to conventional thermodynamic relations
for an ideal gas in a rigid volume.  I also discuss heat capacities for 
trapped particles.
\end{abstract}




\maketitle

\section{Introduction}
The derivation of the ideal gas law is covered in essentially every
modern text on thermodynamics and statistical mechanics.  This law
applies to non-interacting classical particles in a rigid container,
and expresses the familiar relationship
\begin{equation}
PV = NkT
\label{eq_igl}
\end{equation}
between the pressure $P$, volume $V$, number of particles $N$, and the
temperature $T$; $k$ is the Boltzmann constant.  It is also shown that
the energy of of an ideal gas is given by
\begin{equation}
E = \frac{3}{2}NkT.
\label{eq_iel}  
\end{equation}
Recent experiments on cooled neutral atoms (which have ultimately led
to observations of Bose-Einstein condensation) have been performed in
atom traps which in the ideal case have confining potentials of
infinite range.  Thus volume is not an appropriate thermodynamic
variable. It is instructive to consider how the familiar elementary
relations of Eqs.~(\ref{eq_igl}) and (\ref{eq_iel}) must be modified
in these circumstances.  

In this paper I consider atoms in isotropic harmonic confining
potentials.  For particles of mass $m$ and energy $E$ confined by a
one-dimensional harmonic potential with angular oscillation frequency
$\omega$, the amplitude of oscillation is $\sqrt{2E/(m\omega^2)}$.
This suggests using the frequency of oscillation about the minimum of
the trap as the externally determined parameter characterizing the
confinement of the particles in the trap, and in this note I consider
the development of thermodynamic relations which involve $\omega$
rather than the volume $V$.
 
I follow the approach used in several modern texts to derive the ideal
gas law (see, for example, \cite{MAN88,BAI99,SCH00,CAR01}).  The
canonical ensemble partition function is derived in a semi-classical
manner, i.e., using information about the spacing and degeneracy of
quantized energy levels, and the partition function is then used in
combination with the first law of thermodynamics to derive
relationships between the thermodynamic variables.

\section{Fundamental Equations}
\label{sec_sl}
The first law of thermodynamics is an articulation of the work-energy
theorem.  For a gas in a volume $V$ this is expressed as 
\begin{equation}
\Delta E = \Delta Q - P\Delta V.
\label{eq_law1}
\end{equation}
Although the volume isn't a relevant parameter for particles in a trap,
the confining potential can do work on the particles when it is
altered.  We can  express this idea in a modification of the first
the first law,
\begin{equation}
\Delta E = \Delta Q + {\cal P} \Delta \omega,
\label{eq_law1_new}
\end{equation}
in which ${\cal P}$ is a ``pressure'' for which I will derive an
expression below.  (I use the term ``pressure'' loosely; this quantity
does not have the dimensions of force per unit area.)  Note that I
have chosen the sign of the second term on the right side of 
Eq. ~(\ref{eq_law1_new} to be positive, reflecting the fact that 
an increase in $\omega$ corresponds to an increase in the strength of the
confinement, which increases the density of the particles and
effectively compresses them.

The conventional arguments leading to the definition of the 
Helmholtz free energy
\begin{equation}
F= E -TS
\end{equation}
are not affected by the change from confinement in a rigid volume
to confinement in a trap.  Combining the modified first law with
the definition of free energy leads to the relationship
\begin{equation}
dF = -S\,dT + {\cal P}\, d\omega ,
\end{equation}
which implies 
\begin{equation}
{\cal P} = +\left( \frac{\partial F}{\partial \omega}\right)_{T,N}
\end{equation}
and
\begin{equation}
S = -\left( \frac{\partial F}{\partial T}\right)_{\omega,N}.
\end{equation}

The relationship between the Helmholtz free energy and the partition
function is based on general arguments regarding entropy, so that for
particles in a trap it is still true that
\begin{equation}
F(T,\omega,N) = -kT \ln Z(T,\omega,N).
\end{equation}

\section{Explicit Partition Function and Consequences}
\label{sec_pf}
The partition function for a single particle in a a harmonic trap is
\begin{eqnarray}
Z_1 &=& \sum_i e^{-\beta \epsilon_i} \nonumber \\
    &\longrightarrow& \int_0^\infty f(\epsilon)e^{-\beta \epsilon}\, d\epsilon
            \nonumber \\
    &=& \int_0^\infty \frac{\epsilon^2}{2(\hbar\omega)^3} 
	      e^{-\beta \epsilon}\, d\epsilon \nonumber \\
    &=& \left(\frac{kT}{\hbar\omega}\right)^3.
\end{eqnarray}
In the derivation above I have assumed that $kT\gg\hbar\omega$ and
converted the discrete sum into an integral, and I have used the
density of states $f(\epsilon)$ appropriate for a harmonic oscillator
potential.  Using standard arguments, the partition function for $N$
non-interacting particles in a dilute gas is
\begin{eqnarray}
Z &=& \frac{1}{N!}\left[Z_1(T,\omega)\right]^N \nonumber \\
  &\simeq& \left(\frac{e}{N}\right)^N\left(\frac{kT}{\hbar\omega}\right)^{3N}
\end{eqnarray}
where in the last line I have used Stirling's approximation for large $N$.

The Helmholtz free energy is thus 
\begin{eqnarray}
F &=& -kT \ln Z \nonumber \\
  &=& -NkT\ln \left[\frac{e}{N}\left(\frac{kT}{\hbar\omega}\right)^3\right].
\end{eqnarray}
The ``pressure'' ${\cal P}$ is given by 
\begin{eqnarray}
{\cal P} &=& +\left(\frac{\partial F}{\partial \omega}\right)_{T,N}
                   \nonumber \\
         &=& \frac{3NkT}{\omega},
\label{eq_igl_new}
\end{eqnarray}
which gives the analog to the ideal gas law for the trapped particles:
\begin{equation}
{\cal P}\omega = 3NkT.
\end{equation}
The energy of the particles is 
\begin{eqnarray}
E &=& - \left(\frac{\partial \ln Z}{\partial \beta}\right)_{N,\omega}
                \nonumber \\
  &=&  3NkT,
\label{eq_energy}
\end{eqnarray}
which combined with Eqs.~(\ref{eq_igl_new}) shows that
\begin{equation}
E = {\cal P}\omega.
\end{equation}
The linear relationship between energy and oscillation frequency is to
be expected given the fact that energy of each quantized
single-particle energy level is of the form $m\hbar\omega$ (where $m$
is an integer).

\section{Comment on Heat Capacity}
For conventional ideal gases the heat capacity at constant pressure is
\begin{equation}
C_P = \left(\frac{\partial E}{\partial T}\right) = \frac{3}{2}Nk,
\end{equation}
and this is related to the heat capacity at constant volume $C_V$ 
by the familiar equation
\begin{equation}
C_P - C_V = P\left(\frac{\partial V}{\partial T}\right)_P = Nk.
\end{equation}

For trapped particles the heat capacity at constant $\omega$ is
\begin{equation}
C_\omega = \left(\frac{\partial E}{\partial T}\right) = 3Nk.
\end{equation}
This makes sense because the translational motion of each particle
contributes $3k/2$ to the heat capacity, and the potential energy of
the three-dimensional harmonic oscillator confining potential
contributes an additional $3k/2$.  The difference between the heat
capacities at constant $\omega$ and at constant ${\cal P}$ is
\begin{eqnarray}
C_{\cal P} - C_\omega &=& -P\left(\frac{\partial \omega}{\partial T}
                \right)_{\cal P} \nonumber \\
        &=& -3Nk.
\end{eqnarray}
This means that 
\begin{equation}
C_{\cal P} = 0.
\end{equation}
In other words, the energy needed to raise the temperature of the
particles by $\Delta T$ all comes from the work done on the particles
by the increase in $\omega$ that is necessary to keep ${\cal
P}=E/\omega$ constant.

\section{Conclusion}
For particles trapped in smoothly varying long-range potentials, volume
is not an appropriate thermodynamic variable.  Thermodynamics can,
however, be developed using other parameters characterizing the
confinement of the particles.  In this paper I have considered
particles in isotropic harmonic traps, and I have used the frequency
of oscillations about the minimum of the trap as the confinement
parameter analogous to $V$.  I have derived the analog to the ideal
gas law for this simple case, and also formulas for appropriate heat
capacities.  These simple relations can be used to solve many
thermodynamic problems for trapped particles that are analogous to
problems for gases in containers with rigid walls that are posed in
introductory texts.  Generalizations to anisotropic traps and more
complicated potentials are certainly possible, and would make good
student projects.

\end{document}